\documentstyle[prl,aps,amsfonts]{revtex}

\begin{document}

\draft

\title{Evidence for a Second Order Phase Transition in Glasses at Very Low Temperatures --  A Macroscopic Quantum State of Tunneling Systems}

\author{P. Strehlow}
\address{Physikalisch-Technische Bundesanstalt, Abbestrasse 2--12, D-10587 Berlin, Germany}

\author{C. Enss and S. Hunklinger}
\address{
Institut f\"ur Angewandte Physik,Universit\"at Heidelberg,
Albert-Ueberle-Str.\ 3--5, D-69120 Heidelberg, Germany}

\date{\today}

\maketitle

\begin{abstract}
Dielectric measurements at very low temperature indicate that in a glass with the eutectic composition BaO-Al$_2$O$_3$-SiO$_2$ a phase transition occurs at 5.84~mK. Below that temperature small magnetic fields of the order of 10~$\mu$T cause noticeable changes of the dielectric constant although the glass is insensitive to fields up to 20~T above 10~mK. The experimental findings may be interpreted as the signature of the formation of a new phase in which many tunneling systems perform a coherent motion resulting in a macroscopic wave function.
\end{abstract}


\narrowtext

The anomalous low-temperature properties of glasses are caused by low energy excitations present in all amorphous solids\cite{Top84}. In the tunneling model~(TM)\cite{Phi72,And72} these excitations are described on a phenomenological basis by tunneling systems (TSs). In the simplest case, such a tunneling system can be represented by a particle in a double-well potential. At sufficiently low temperature only the ground states in the two wells are relevant, and the system will effectively be restricted to the two-dimensional Hilbert space spanned by the two ground states. In the pseudo-spin representation with the Pauli matrices $\bbox{\bbox{\sigma}}^x$ and $\bbox{\bbox{\sigma}}^z$, the Hamiltonian of an isolated TS is given by $H_0 =(1/2)({\it\Delta}\bbox{\bbox{\sigma}}^z - {\it\Delta}_0\bbox{\sigma}^x)$, where ${\it\Delta}$ and ${\it\Delta}_0$ are the asymmetry and the tunneling matrix element, respectively. Because of the random structure of glasses the characteristic parameters~${\it\Delta}$ and ${\it\Delta}_0$ of the TSs exhibit a broad distribution. Clearly, the Hamiltonian of an isolated TS is formally equivalent to that of a spin~1/2-particle in a magnetic field. This analogy can advantageously be used to describe the dynamics of TSs in glasses. The interaction of the TSs with acoustic and electric fields is usually treated as a weak perturbation of $H_0$. In this way the standard TM successfully explains many of the anomalous thermal, acoustic and dielectric properties of glasses at temperatures below 1~K.

Upon closer consideration, however, pronounced deviations from the expected behavior are found in many cases. They probably have their origin in the interaction between the TSs which is not taken into account in the standard TM. The strength of this interaction was first deduced from acoustic hole burning experiments\cite{Arn75} and studied in more detail by measurements of the decay of coherent echoes\cite{Gra79}. Moreover, memory effects have been observed in dielectric measurements\cite{Rog96} which indicate the influence of the long-range interaction of TSs on both, the density of states and the dielectric response\cite{Bur95}. Very recently, it has been argued that a transition from coherent to incoherent tunneling takes place if the mean interaction energy exceeds the tunnel splitting\cite{Ens97}. Taking into account the peculiarities of glasses, this concept allows a quantitative explanation of several inconsistencies arising in the standard TM.

For lack of knowledge of the microscopic nature of the TSs, the interaction between them has to date been described in terms of an elastic (or electrostatic) coupling mediated by virtual phonon (or photon) exchange. Based on the spin-boson Hamiltonian\cite{Leg87}, the elastic interaction between the TSs in glasses can be investigated in a non-perturbative manner\cite{Kas89}. These calculations demonstrate that the strong TS-phonon coupling essentially leads to a renormalization of the tunneling parameter and thus modifies related quantities like density of states or relaxation times. This result explains to some extent the success of the standard TM: differently strong coupling between the TSs does not cause qualitative but only quantitative changes. However, it is known from the widely used spin~1/2-Ising model that a variation of the interaction may alter the macroscopic properties of an ensemble of two-state systems even qualitatively. This means that the exchange Hamiltonian $H_{ \rm int} = - (1/2)\sum_{i,j} J_{ij}\bbox{\sigma}_i^z\bbox{\sigma}_j^z$, with the potential~$J_{ij}$ between TSs at sites~$i$ and~$j$, is responsible for the occurrence of cooperative processes and even of continuous phase transitions.

In fact, indications of the existence of a phase transition were found for the first time in multicomponent glasses. In measurements below 10~mK an abrupt increase of the sensitivity of the dielectric constant to magnetic fields was observed. The dielectric properties of these glasses have intensively been investigated\cite{Stre94} because of their applicability to capacitance thermometry under high magnetic fields. In particular, it has been established that at temperatures down to 16~mK magnetic fields up to 20~T have an imperceptible effect on the dielectric constant\cite{Pen95}. In order to confirm the experimental evidence for the occurrence of a phase transition in glasses we have studied the dielectric properties of a glass with the eutectic composition BaO-Al$_2$O$_3$-SiO$_2$. As sample we used a thick-film capacitance sensor ($10\times 10 \times 0.05~{\rm mm}^3$) with 30~$\mu$m thick gold electrodes on a sapphire substrate. The sensor was prepared from glass powder and gold paste by the silk-screen process and subsequent sintering at 1225~K\cite{Heraeus}.

We report here on measurements of the dielectric constant (or permittivity) at temperatures between 700~$\mu$K and 1~K. The experiments were carried out in a nuclear demagnetization cryostat using an automatically balancing capacitance bridge operating at 1~kHz. In order to avoid uncontrolled variations of the magnetic field due to the stray field of the magnet, the sample was placed in a Nb-cylinder with 25~mm in diameter. Although efforts were made to screen the earth magnetic field and that of the magnet, a residual field of $B_0 \approx 20~\mu$T was present at the sample. Passing the transition temperature of Nb this field is frozen in. To allow for a variation of the magnetic field the sample was mounted inside a small coil fixed within the cylinder. Based on an extended temperature scale\cite{Schu96} the temperature was measured with a $^3$He melting curve thermometer (MCT) and a pulsed platinum NMR thermometer.

In Fig.~1 we have plotted the temperature variation $\delta \varepsilon'(T) / \varepsilon'(T_0) = [\varepsilon'(T) - \varepsilon'(T_0)]/\varepsilon'(T_0)$ of the dielectric constant~$\varepsilon'$, where $T_0 = 1.26$~mK is the temperature of the lowest lying data point. At high temperature the permittivity decreases with decreasing temperature, passes a minimum around 110~mK and increases again. It levels off at a few Millikelvin and approaches a constant value at the lowest temperature. An analogous behavior has been reported for other glasses\cite{Nish92}.

Starting from higher temperature, the dielectric constant decreases due to relaxation processes which vanish with decreasing temperature. The increase of $\varepsilon'$ below the minimum reflects the resonant interaction of the TSs with the electric field. The prediction of the standard TM is shown in Fig.~1 by a dashed line.
At high temperature the agreement between fit and data is satisfactory whereas at low temperature significant deviations occur.
As mentioned above, the variation of $\varepsilon'(T)$ in the entire temperature range can be understood within the frame of the TM if the occurrence of incoherent tunneling is taken into account\cite{Ens97}. As in other glasses\cite{Nish92,Stre96} the exact change of the dielectric constant at low temperature depends on the strength of the applied electric field which was 120~kV/m in the measurement shown.

Although the occurrence of incoherent tunneling and the field dependent dielectric response are interesting effects in themselves, we not discuss them here but turn to even more exciting results. We have carried out a very careful measurement of $\varepsilon '$ at temperatures around 5~mK. The sample was mounted in the Nb-cylinder to keep the effective magnetic field constant and was slowly cooled from 6.88~mK to 4.88~mK at a constant rate of 62.6~$\mu$K/min. In the upper part of Fig.~2 the recorded variation of the dielectric constant is plotted as a function of time. Within the narrow temperature range of 2~mK it can be approximated by two straight lines crossing at 5.84~mK. In order to magnify this discontinuity we have plotted in the lower part of the figure the deviation~$\delta\varepsilon'_{\rm s}$ of the dielectric constant from the straight line through the first and last data point. This plot demonstrates that there is an abrupt change of the slope indicating a
phase transition at 5.84~mK.

Further extremely surprising results were obtained in studies of the influence of magnetic fields onto the dielectric constant which suddenly set in below 5.84~mK. In Fig.~3 the relative change $\delta \varepsilon'/\varepsilon'$ of the dielectric constant at 1.85~mK in the presence of a time dependent magnetic field is plotted as a function of time. At the top of the figure the variation~$\delta B(t) = B(t) - B_0$ of the field is shown. The baseline in this plot is determined by the frozen-in field~$B_0 = 20~\mu$T. In the lower part of the figure the data on $\delta\varepsilon' /\varepsilon'$ are presented. Each data point was obtained by averaging the reading of the bridge for one minute. Note that the scatter of $\delta \varepsilon' / \varepsilon'$ is only of the order of $10^{-8}$. The surprising result is that the dielectric constant follows the variation of the magnetic field.

In Fig.~4 the same type of measurement is shown for five different temperatures. As can be seen at the top of this figure the field changes now had the opposite sign. Astonishingly, with decreasing temperature the variation~$\delta \varepsilon'$ associated with $\delta B$ completely changes character. At 11~mK the dielectric constant is independent of the applied field within the scatter of our data. At 5.07~mK small changes can be seen which become clearly visible at lower temperature. They are most pronounced around 2~mK. Below that temperature the influence of the field diminishes and finally at 0.72~mK the dielectric constant is no longer influenced by these small fields. Taking a look at the noise level reveals another surprising phenomenon. In going from 11~mK to 5.07~mK the scatter of the data at zero-field is drastically reduced and remains small down to the lowest temperature.

There is an important difference between the data taken at 1.40~mK and 2.56~mK, and those shown in Fig.~3. The dielectric constant first follows the variation of the magnetic field. However, the decrease of $\varepsilon'$ does not stop when the zero-field~$B_0$ is reached. It continues to decrease for several minutes and finally approaches the equilibrium value again. This behavior can be understood if the change of $\varepsilon'$ caused by the magnetic field is accompanied by the production of heat. In this case the sample warms up and $\varepsilon'$ decreases according to Fig.~1. Since the heat is slowly flowing to the heat sink the temperature tends to its lower equilibrium value and the dielectric constant recovers with time.

Summarizing we may state that we have observed two surprising phenomena: The sharp discontinuity exhibited by the dielectric constant resembles strongly the ``A - feature" observed in the melting curve of $^3$He and indicates the existence of a hitherto unknown second order phase transition in glasses. The magnetic field dependence of the dielectric permittivity is rather astonishing because no first order dependence between the dielectric constant and the magnetic field should exist.

In an isotropic dielectric, devoid of free charge and free current, the general constitutive equation for the polarization ${\bf P}$ is of the form $\bbox{P} = \varepsilon_0\chi_1 \bbox{\frak E} + \varepsilon_0\chi_2 (\bbox{B}\cdot \bbox{\frak E})\bbox{B}$, where the coefficients $\chi_1$ and $\chi_2$ may depend on temperature, density and the three scalars ${\frak E}^2$, $B^2$, and $(\bbox{\frak E}\cdot \bbox{B})^2$. Polarization and electromotive intensity~$\bbox{\frak E} = \bbox{P} + \bbox{v} \times \bbox{B}$ are polar vectors, while the magnetic induction~$\bbox{B}$ is an axial vector. This means that under inversion, $\bbox{P}$ and $\bbox{\frak E}$ change their signs, while $\bbox{B}$ remains invariant. Glasses are properly assumed to be linear dielectrics so that the polarization is given by $\bbox{P} = \varepsilon_0\chi_1 \bbox{\frak E} = (\varepsilon - \varepsilon_0) \bbox{\frak E}$ and the dielectric constant~$\varepsilon$ depends only on temperature and density (or deformation). Deviations from a linear behavior require a nonlinear constitutive equation for the magnetization~$\bbox{M}$ because of the integrability condition $\partial \bbox{P} / \partial \bbox{B} = \partial \bbox{M} / \partial \bbox{\frak E}$. Although the observed dependence of $\varepsilon$ on $\bbox{B}$ can formally be explained in a nonlinear field theory, the abrupt appearance of a nonlinear magnetic behavior in glasses below 6~mK remains to be seen. It is tempting to discuss the discovered low temperature phenomenon as a continuous phase transition with a critical temperature~$T_{\rm c}$. For $T< T_{\rm c}$, the thermal noise is too weak to suppress the formation of a highly coherent ensemble of TSs dressed with clouds of virtual phonons. For energies smaller than the characteristic energy scale given by $k_{\rm B} T$, the dressed TSs may be considered as elementary excitations of a ground state described by the macroscopic wave function ${\it\Psi} = \vert {\it\Psi}\vert \exp ({\rm i}{\it\Phi})$. Since the equilibrium value of the modulus of ${ \it\Psi}$ is determined by the temperature, the arbitrary phase~${\it\Phi}$ parametrizes a continuously degenerated equilibrium state of glasses below~$T_{\rm c}$. The distinguishability of the phase implies that the gauge symmetry~U(1) is spontaneously broken. The inhomogeneous phase structure occurring below $T_{\rm c}$ can be studied by means of the Ginzburg-Landau functional. Due to the coupling between $\bbox{B}$ and ${\it\Psi}$ the influence of the magnetic field on the polarization below $T_{\rm c}$ can qualitatively be described.

In conclusion we have reported on surprising phenomena discovered in BaO-Al$_2$O$_3$-SiO$_2$-glass at very low temperatures which have never been observed so far. We want to point out that we reproduced the measurements not only in the case of BaO-Al$_2$O$_3$-SiO$_2$-glass but have made similar findings also in other multicomponent glasses though $T_{\rm c}$ exhibited slightly different values. Therefore, we expect that the observed phase transition from uncorrelated, incoherent tunneling of individual TSs to a correlated motion of a large number of dressed TSs occurs in all glasses. Of course, we have only touched on a novel and unexpected phenomenon and much more experimental and theoretical work is need to fully understand the underlying physical principles.

The authors would like to thank G. Schuster, A. Hoffmann, and D. Hechtfischer for putting the MCT data to our disposal and for helping us with to the realization the nuclear demagnetization experiments. Furthermore, we are grateful to Ch. Modes and K. Deckelmann of Heraeus GmbH for preparing the thick-film sensors and acknowledge discussions with J. Classen and M. von Schickfus. This work has been supported by the Deutsche Forschungsgemeinschaft (Grant: Hu$\,$359/11).
%


\begin{figure}
\caption{Temperature variation of the dielectric constant~$\varepsilon '$ of BaO-Al$_2$O$_3$-SiO$_2$-glass measured at 1~kHz. The dashed line represents a calculation with the tunneling model.}
\label{dielectric constant}
\end{figure}
\begin{figure}
\caption{
a) Dielectric constant of the BaO-Al$_2$O$_3$-SiO$_2$-glass measured at a constant cooling rate of 62.6~$\mu$K/min. The dashed line is the extrapolation of the data from higher temperature. b) Same data after subtracting a straight line through the first and last point of the measurement. On the abscissa the time scale has been converted to temperature.}
\label{phase transition}
\end{figure}
\begin{figure}
\caption{
Influence of the magnetic field on the dielectric constant of the Ba0-Al$_2$0$_3$-SiO$_2$-glass. a) Time variation~$\delta B(t)$ of the applied magnetic field. b) Relative change of the dielectric constant with the variation of the applied magnetic field at 1.85~mK.}
\label{magnetic field down}
\end{figure}

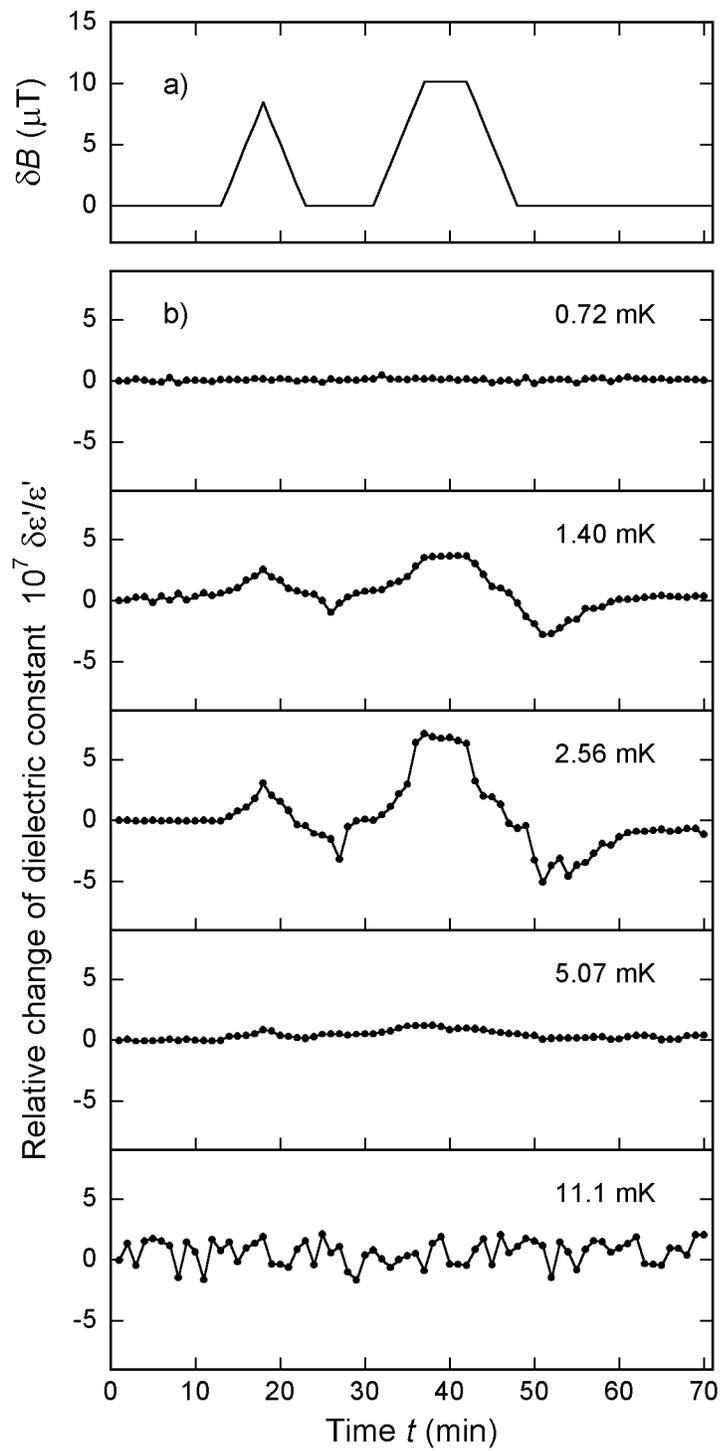
\begin{figure}
\caption{
Influence of the magnetic field on the dielectric constant of the BaO-Al$_2$O$_3$-SiO$_2$-glass. a) Time variation~$\delta B(t)$ of the applied magnetic field. b) Relative change of the dielectric constant with the variation of the applied magnetic field at temperatures between 0.72~mK and 11.1~mK.}
\label{magnetic field up}
\end{figure}


\begin{references}

\bibitem{Top84}   {{\it Amorphous Solids}, Topics in Current Physics {\bf
                  24}, edited by W.A. Phillips (Springer, Berlin, 1984).}

\bibitem{Phi72}   {W.A. Phillips, J. Low Temp. Phys. {\bf 7}, 351 (1972).}

\bibitem{And72}   {P.W. Anderson, B.I. Halperin, and C.M. Varma, Philos.    Mag. {\bf 25}, 1 (1972).}

\bibitem{Arn75}   W. Arnold and S. Hunklinger, Solid State Commun.
                  {\bf 17}, 833 (1975).

\bibitem{Gra79}   J.E. Graebner and B. Golding,
                  Phys. Rev. B {\bf 19}, 964 (1979).

\bibitem{Rog96}   S. Rogge, D. Natelson, and D.D. Osheroff, Phys. Rev. Lett. {\bf 76}, 3136 (1996).

\bibitem{Bur95}   A.L. Burin, J. Low Temp. Phys. {\bf 100}, 309 (1995).

\bibitem{Ens97}   C. Enss and S. Hunklinger, Phys. Rev. Lett. {\bf 79}, 2831 (1997).

\bibitem{Leg87}   A.J. Leggett, S. Chakravarty, A.T. Dorsey, M.P.A. Fisher, A. Garg, and W. Zwerger, Rev. Mod. Phys. {\bf 59}, 1 (1987).

\bibitem{Kas89}   K. Kassner and R.J. Silbey, J. Phys.: Condens, Matt. {\bf 1}, 4599 (1989).

\bibitem{Stre94}   P. Strehlow, Cryogenics {\bf 34}, 421 (1994).

\bibitem{Pen95}   F. Penning, M. Maior, P. Strehlow, S. Wiegers, H. van Kempen, and J. Maan, Physica B {\bf 211}, 363 (1995).

\bibitem{Heraeus}  The thick-film sensors were manufactured by W.C. Heraeus GmbH, thick-film technology, Heraeusstrasse 12-14, D-63450 Hanau, Germany.

\bibitem{Schu96}   G. Schuster, A. Hoffmann, and D. Hechtfischer, {\it Temperature scale extension below ITS~90 based on $^3$He melting pressure}, CCT -- 96/25 (1996). 

\bibitem{Nish92}   H. Nishiyama, H. Akimoto, Y. Okuda, and H. Ishimoto, J. Low Temp. Phys. {\bf 89}, 727 (1992).

\bibitem{Stre96}   P. Strehlow, C. Enss, and S. Hunklinger, Czech. J. Phys.
                  {\bf 46}, 2231 (1996).

\end{references}
\end{document}